\documentclass[twoside,10pt,a4paper]{newFNLstyle}
\usepackage{graphics}
\usepackage{cite}

\begin{document}

\volnumpagesyear{4}{3}{L?}{2004}
\dates{8 June 2004}{30 July 2004}{5 August 2004}

\title{Stochastic phase-field simulations of symmetric alloy solidification}

\authorsone{R. Ben\'{\i}tez and L. Ram\'{\i}rez-Piscina}
\affiliationone{Department de F\'{\i}sica Aplicada,
Universitat Polit\`ecnica de Catalunya}
\mailingone{Doctor Mara\~n\'on 44, E-08028 Barcelona, Spain.}
  


\maketitle

\markboth{Stochastic Phase-Field Simulations of Symmetric Alloy Solidification}{R. Ben\'{\i}tez and 
L. Ram\'{\i}rez-Piscina}

\pagestyle{myheadings}

\keywords{Phase-field models; fluctuations; solidification; transient stages.}

\begin{abstract} 
We study initial transient stages in directional solidification 
by means of a non-variational phase field model with fluctuations. 
This model applies for the symmetric solidification of dilute binary solutions 
and does not invoke fluctuation-dissipation theorem to account for the
fluctuation statistics.
We devote our attention to the transient regime during which concentration
gradients are building up and fluctuations act to destabilize the interface. To 
this end, we calculate both the temporally dependent growth rate of each mode and the
power spectrum of the interface evolving under the effect of fluctuations.
Quantitative 
agreement is found when comparing the phase-field simulations with theoretical 
predictions.
\end{abstract}

\section{Introduction}
\label{sec:1}

Solidification processes constitute a primary subject of research in materials
science. 
One main objective
is to achieve the prediction and control of the final
macroscopic properties (mechanical, electrical, etc) of solids grown from their
melts.
These properties largely 
depend on the microstructure, 
which in turn is result of the 
conditions at which the solid was grown \cite{annualreview}. 
For example, relevant features at the microscopic scale such as compositional
inhomogeneities or the presence of grain boundaries are direct consequences of
instabilities and subsequent dynamics of the solidification front.
Of particular interest for the metallurgical community has long been the
directional solidification of mixtures and alloys, of direct relevance in
processes of zone melting and for the Bridgman method \cite{kurz}. 
This configuration is also an archetypical model system in nonlinear physics to
study pattern selection and complex dynamics \cite{langerrmp}. 
In this context it is known that the final wavelength of the dendritic array
depends on the history \cite{wl-1}, and specifically on the initial destabilization
of the solidification front \cite{wl-2,laure-caroli}. 
This is a non-steady noise amplification process occurring during the early solute
redistribution transients, for which considering both fluctuations and transient
effects are of key importance. Experimental work in this regime can be found 
in \cite{figue1,figue2}. We are interested here in dealing with
these aspects in numerical simulations of a directional solidification experiment,
by using a phase field model supplemented with fluctuations.

During the last decade, the phase-field model has emerged 
as a quantitative simulation technique to study complex interfacial 
morphologies \cite{CincaBook04}. Recently, the phase field approach 
has deserved much attention because it can easily incorporate effects like 
system anisotropies, kinetic attachment or equilibrium 
fluctuations which might be difficult to include with other 
simulation techniques. This method avoids the tracking of the 
moving front by introducing a continuous order 
parameter $\phi(\vec{r},t)$ (the phase field),
which takes different constant values at the bulk phases 
separated by a smooth interface of width $W$. 
The model then consists in a set of coupled equations for $\phi(\vec{r},t)$ 
and for the diffusion field which drives the interfacial dynamics, and that are
constructed so that it reproduces the physical dynamics in the limit $W \rightarrow 0$.
Early formulations of phase field models were variational, i.e. the model
equations were derived from functional derivatives of a single free energy
functional, but the convergence of such models was rather poor.
Recent advances in phase field formulations, either based in the so called thin
interface limit \cite{Karma96,Karma01} or in higher order expansions \cite{almgren-one-sided},
permitted to improve dramatically the convergence of the model, while 
dealing with some specific realistic situations.

However the standard procedure to introduce fluctuations in phase field models is
restricted to 
variational formulations \cite{Karma99,Pavlik99,Pavlik00}, analogous to the model
C of critical dynamics
\cite{Halperin74,wheeler1,wheeler2}, where the intensity of the fluctuations 
can be determined by using a fluctuation-dissipation relation.  
As the more recent phase field formulations do not maintain this variational
structure, the fluctuation-dissipation relation 
cannot be used to infer the statistics of the noise appearing in 
the equations. In the model and simulations presented here, we 
rely on a recent calculus \cite{benitez} that projects 
the dynamics of a generic stochastic phase field model to the motion of the 
fluctuating interface. This procedure provides a prescription for the 
intensity of the noise terms in the model, accounting for fluctuations of both 
internal and external origin. This approach has previously been used in
equilibrium situations only \cite{benitez}, and our aim here is to employ it in
a transient, out of equilibrium, situation.
We will restrict ourselves to the study of the initial destabilization in the
linear regime, where comparison with theory is possible, as a first step to the
complete problem of considering the whole competition process in the presence of
fluctuations.
The validation of this approach should be also useful in other situations of
interfacial dynamics with fluctuations, both in non variational formulations of
phase field models and in considering more general sources of fluctuations, of
external origin.

\section{Stochastic Phase-Field Model for Symmetric Directional Solidification}

  In a directional solidification experiment, the mixture sample is pulled at a
velocity $\tilde{v}_p$ in an externally imposed temperature gradient given by 
$T(\tilde{z}) = T_M + \tilde{G} \tilde{z}$, where $T_M$ is the melting temperature
of the mixture and the tilde refers to physical units.
For pulling velocities higher than a certain critical value, the interface becomes
unstable \cite{mullins-sekerka} giving rise to cellular regimes. 
The diffusive field which drives the 
interface dynamics is the reduced solute concentration field 
$u=\frac{C-C_\infty}{\Delta C_0}$,  
where $C_\infty$ is the solute concentration of the sample 
in the liquid bulk far from the solid-liquid interface 
and $\Delta C_0= C_L^0-C_S^0$ is the miscibility gap 
which is approximated to be constant in this model. 
We are interested in the symmetric case for which the 
diffusion coefficients in both phases are equal, situation which is typical in 
dilute mixtures of liquid crystals.
We consider a modified version of 
a non-variational phase-field model presented by Karma \cite{Karma98,Karma01} 
for the directional solidification of a symmetric dilute alloy. 
Scaling space and time with $l=D/\tilde{v}_p$, $\gamma=l^2/D$, being $D$ the 
solutal diffusivity, we obtain the phase-field equations
\begin{eqnarray}
\alpha \varepsilon^2 \partial_{t} \phi &=& \varepsilon^2\nabla^2\phi-
f'(\phi)- \varepsilon \lambda  g'(\phi) (u+\frac{z-t}{l_T}) + 
\varepsilon^{\frac{3}{2}}\eta({\bf r},t)
\label{eq:pf-2}\\
\partial_{t} u &=& \nabla^{2}u+\frac{1}{2}\partial_{t} h(\phi) -
{\bf \nabla}\cdot {\bf q}({\bf r},t)\;,
\label{eq:u-2} 
\end{eqnarray}
with
\begin{eqnarray}
\langle
\eta({\bf r},t) \eta({\bf r'},t') 
\rangle &=&
2 \sigma_\phi^2 \delta({\bf r}-{\bf r'}) 
\delta(t-t')
\label{eq:eta-2}\\
\langle
q_i({\bf r},t)
q_j({\bf r'},t')
\rangle &=&
2 \sigma_u^2 \delta_{ij} \delta({\bf r}-{\bf r'})\delta(t-t')\;.
\label{eq:q-2} 
\end{eqnarray}
In the last equations, $\varepsilon=W/l$ stands for the dimensionless 
thickness of the interface, and $l_T=m \Delta C_0/G$ is the thermal length imposed
by the 
external gradient, being $m$ the liquidus slope of the alloy. 
Fluctuations have been included following the procedure 
introduced in \cite{benitez} to account for the equilibrium statistics. 
It can be shown that the 
model defined by Eqs.~(\ref{eq:pf-2}), (\ref{eq:u-2}), (\ref{eq:eta-2}) and (\ref{eq:q-2}) 
recovers the classical moving boundary description of the problem 
in the limit of $\varepsilon \rightarrow 0$ (small interface width) 
and $\sigma_\phi,\sigma_u \ll 1$ (small noise intensities). 
For this particular model, and performing a matching asymptotic 
procedure up to first order in $\varepsilon$, we find the relations 
$\lambda = \frac{I_1}{I_2}\frac{1}{d_0}$,  
$\alpha = \frac{\beta + I_4\varepsilon}{d_0}$, 
$\sigma_\phi^2 = \frac{\beta v_m I_1}{d_0^2 \Delta C_0 (1-k) N_A l^d}$, and 
$\sigma^2_u = \frac{C^0_L v_m}{N_A \Delta C_0^2 l^d}$, 
where $I_1$, $I_2$ and $I_4$ are integral constants 
given by $I_1=2\sqrt{2}/3$, $I_2=16/15$ and $I_4=0.55$, 
and we have chosen the particular set of functions 
$f(\phi) = -\frac{1}{2}\phi^2 + \frac{1}{4}\phi^4$, 
$g(\phi) = \phi-   \frac{2}{3}\phi^3+\frac{1}{5}\phi^5$  
and $h(\phi) = \phi$. In the last relations, $\beta$ is the 
kinetic attachment, $k$ the segregation coefficient, $N_A$ the Avogadro's number, 
$v_m$ the molar volume and $d_0=\sigma T_M / L m \Delta C_0$ the chemical
capillary length, 
where $\sigma$ is the surface energy and $L$ the latent heat per unit volume.  
\section{Simulation Results} 
\label{sec:4}
As a quantitative benchmark of the results presented in \cite{benitez}, 
we perform numerical simulations of the 
phase-field model and compare them with theoretical predictions obtained from the moving boundary 
description of the front. 
We will present results for both the stationary power spectrum of the interfacial 
fluctuations and for the wavelength selection during the initial recoil transient. 

The substance parameters correspond to the liquid crystal 
4-n-octylcyanobiphenil (8CB), which has been used in several quantitative 
experimental works \cite{oswald,simon,figue1}. 
The substance parameters have been obtained from \cite{oswald,simon,figue1}, and 
are given by $k=0.9$, $m=88.46$ K/mol, 
$T_M=313.5$ K, $\sigma=2.2\times 10^{-4}$ J/m$^2$, 
$\beta=113.04$ s mol/m,  $D =4\times 10^{-10}$ m$^2$/s,  
$L=2.2\times 10^{6}$ J/m$^3$, 
$Z=291.44$ g (molecular weight) and 
$\rho=10^6$ g/m$^3$ (density). The experimental parameters chosen in the
simulations 
are $C_\infty=0.012$ mol, $\tilde{v}_p=6\times10^{-5} m/s$ and 
$\tilde{G}=2.3\times10^{3} K/m$.
%
\subsection{Stationary power spectrum} 
\label{sec:4.2}
\begin{figure}[t]
\vspace{0.2cm}
\centering{\resizebox{5cm}{!}{\includegraphics{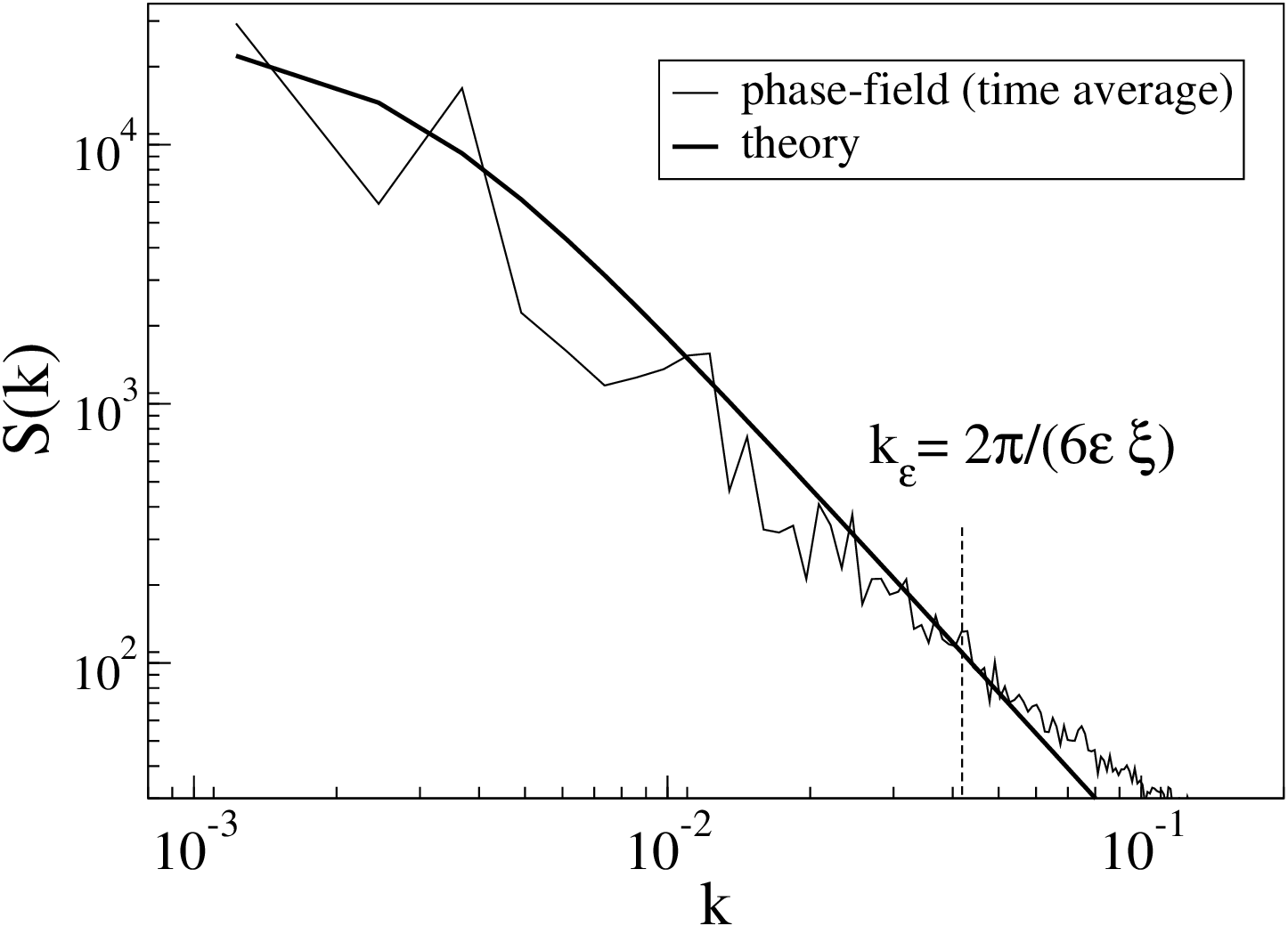}}}
\caption{Mean-square interfacial fluctuation spectrum 
obtained from phase-field simulations and compared with 
the theoretical prediction.} 
\label{fig:zz}
\end{figure}

We have simulated a planar, stationary interface in order to 
evaluate its spectral statistical properties. In particular, we consider 
the mean-square fluctuation spectrum, which can be predicted analytically by 
$S(k)= 
\int \frac{dk'}{2\pi} \langle z_f(k) z_f(k') \rangle=
\frac{m v_m}{N_A (1-k) (G + md_0k^2)}$, 
being $z_f(k,t)=\int dk \; z_f({\bf r},t) 
\; e^{-i {\bf k} {\bf r}}$ the Fourier transform of the interface 
position.
Figure \ref{fig:zz} compares the theoretical power spectrum $S(k)$ with the 
power spectrum obtained by analyzing the spectral properties 
of the phase-field interface. This simulation has been performed by using an 
explicit finite differences scheme in a $N_x \times N_z=256 \times 100$
rectangular grid with a scaled time step of $\Delta t=6.75 \times 10^{-5}$ 
and a scaled grid spacing $\Delta z=\Delta x=0.03$. 
The interface width is $\varepsilon=0.0375$, and the phase-field 
simulations have been thermalized during $10^4$ time steps 
before calculating any of the magnitudes presented in this work. 
The interfacial fluctuation spectrum has been averaged among $10^4$ samples 
taken every 10 simulation steps in order to avoid statistical time correlations.
The quantitative agreement indicates the correctness of the determination of 
the model parameters. The scale associated to the 
interfacial width $\varepsilon$, which constitutes the limit 
of the descriptions of any phase-field model, is indicated by a vertical dashed
line 
and determines the maximum wavelength for which the simulations results are 
reliable.
\subsection{Transient power spectrum and wavelength selection}
\label{sec:4.3}
\begin{figure}[t]
\vspace{0.26cm}
\centering{\resizebox{6cm}{!}{\includegraphics{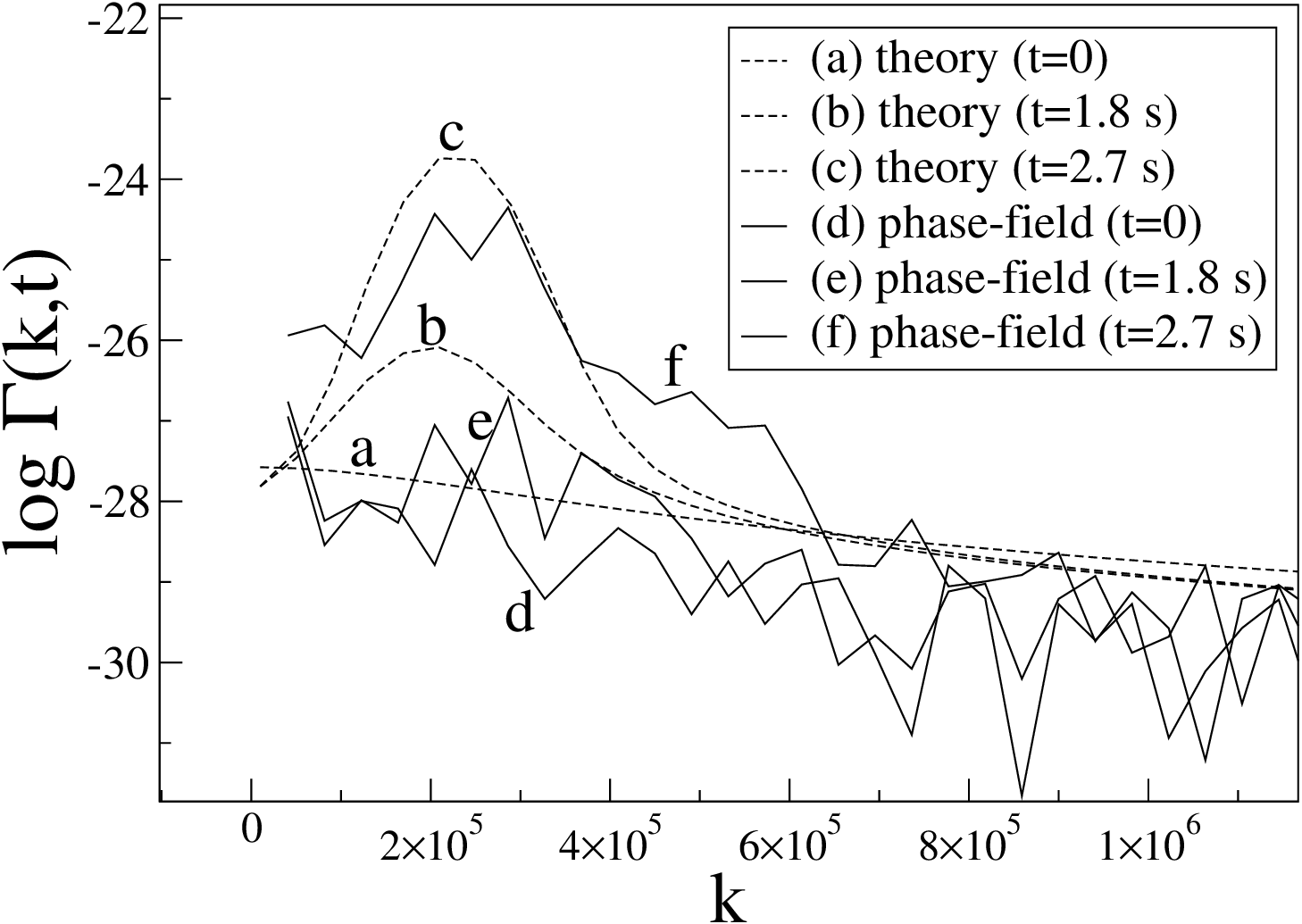}}}
\caption{Noise amplification and wavelength selection 
during the transient. Comparison between the theoretical prediction 
and phase-field simulation results.} 
\label{fig:spec}
\end{figure}

\begin{figure}[htbp]
\vspace{0.1cm}
\centering{\resizebox{5cm}{!}{\includegraphics{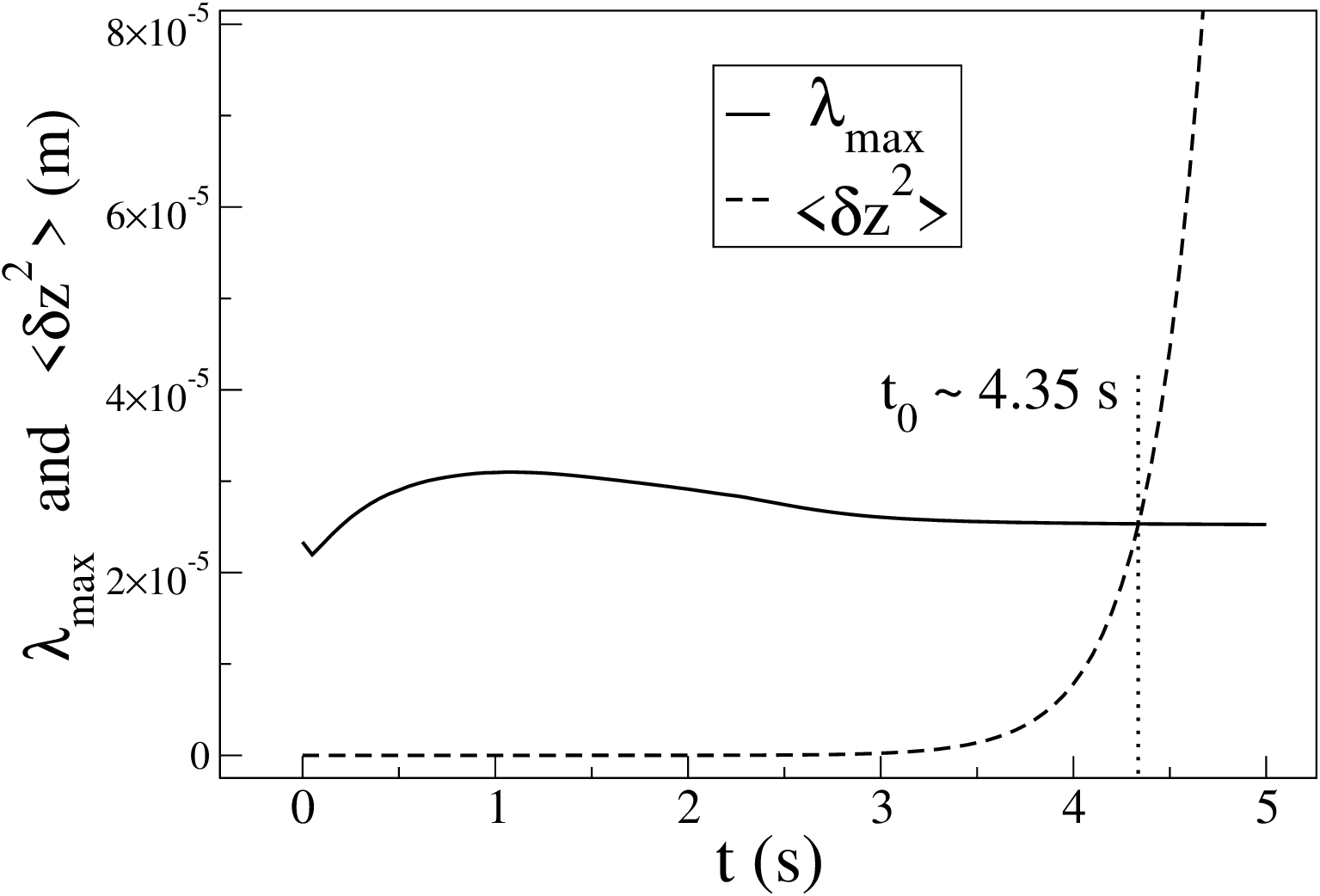}}}
\caption{Determination of the crossover time $\tilde{t}_0$.} 
\label{fig:cross}
\end{figure}

We next investigate the amplification of the 
fluctuations during the initial redistribution transient. 
The simulations are compared with a theoretical 
prediction based on the Warren and Langer's work on 
noise amplification~\cite{wl-2}. 
In this reference, the time evolution of the interface correlations 
is described by $S(k,t) = S_0(k) \int_{-\infty}^{t} dt' 
\exp \left\{2 \int_{t'}^{t} \omega (k,t) dt' \right\}$ 
where $\omega(k,t)$ is the transient growth rate 
of the symmetric model \cite{dresden}, and $S_0(k)$ is given by 
$S_0(k)=\frac{|\omega^e| K_B T^2_M}
{LG(1+d^c_0 \;l_T \;k^2)}$,
being $\omega^e=\omega(k,t=\infty)$ the amplification 
rate at $t \rightarrow \infty$. 
Figure \ref{fig:spec} shows the amplification of the 
interfacial fluctuation spectrum at three different physical times 
during the initial transient ($\tilde{t}$= 0, $1.8$ s and $2.7$ s). 
Dashed lines are obtained by integrating $S(k,t)$ 
in the adiabatic regime $\omega \ll k^2$.
In this case the interface thickness is $\varepsilon=0.15$, and 
the simulations have been performed in a $500 \times 256$ grid 
with $\Delta z=\Delta x=0.09$, $\Delta t=0.00135$. The resulting power spectrum 
is noisy because the results are direct spectral properties 
of the fluctuating interface at each time and no 
average or filtering has been used.
As it can be observed, quantitative agreement is achieved between theory 
and simulations in the early time amplification of fluctuations as well as 
in the wavelength selection of the cellular pattern.
After the early stages where the growth of fluctuations is linear, 
the system enters in a nonlinear regime and a cellular pattern with a 
certain wavelength sets in the system. The end of the 
linear regime can be defined as the time at which the amplitude 
of the most unstable Fourier mode is comparable with its wavelength. 
This condition defines a crossover time $t_0$ which can be
theoretically determined by the condition 
$<\delta z^2> \sim \lambda_{max}$, where 
$<\delta z^2> = \int \frac{d^2 k}{(2\pi)^2} S(k,t)$ 
is the mean-square fluctuation amplitude and $\lambda_{max}$ is the 
wavelength of the largest Fourier mode. Figure \ref{fig:cross} 
shows the evolution in time of these two magnitudes in our system, and 
indicates the determination of the crossover time $\tilde{t}_0$ in physical units.
%
%
 
\section{Conclusions}
We have studied the initial transients of a directional
solidification experiment of a mixture of liquid crystals in the
context of the symmetric model with constant concentration gap. 
Contrary to previous work on transients in fluctuating phase-field
simulations \cite{dresden}, results presented here have been
obtained by using a non-variational formulation of the stochastic 
phase-field model, recently proposed \cite{benitez}. 
Fluctuations have been implemented giving the correct 
equilibrium statistics but without employing any 
fluctuation-dissipation relation to determine the noise intensities in 
the model. The agreement between predictions and simulations has been excellent,
thus confirming the applicability of this approach for the study of 
fluctuations during transient stages. As a final conclusion, the employ of this
class
of efficient stochastic phase-field models opens
new possibilities for the quantitative study by simulation of
microstructure formation in solidification
processes.
%
\section*{Acknowledgements}  
This work was financially supported by
Direcci\'on General de Investigaci\'on Cient\'{\i}fica y T\'ecnica
(Spain) (Project BFM2003-07850-C03-02) and
Comissionat per a Universitats i Recerca (Spain) (Project
2001/SGR/00221). 
%
\bibliographystyle{FNL2}
\bibliography{benbib}

\begin{thebibliography}{10}

\bibitem{annualreview}
W.~J. Boettinger, S.~R. Coriell, A.~L. Geer, A.~Karma, W.~Kurz, M.~Rappaz and
  R.~Trivedi {\em Acta. Mater.\/} {\bf 48} (2000) 43.

\bibitem{kurz}
W.~Kurz and D.~Fisher {\em Fundamentals of Solidification\/} (Trans Tech Pub.,
  1989).

\bibitem{langerrmp}
J.~Langer {\em Rev. Mod. Phys.\/} {\bf 52} (1980) 1.

\bibitem{wl-1}
J.~A. Warren and J.~S. Langer {\em Phys. Rev. A\/} {\bf 42}, 6 (1990)
  3518--3525.

\bibitem{wl-2}
J.~A. Warren and J.~S. Langer {\em Phys. Rev. A\/} {\bf 47}, 4 (1993)
  2702--2712.

\bibitem{laure-caroli}
B.~Caroli, C.~Caroli and L.~Ram\'{\i}rez-Piscina {\em J. Crys. Growth\/} {\bf
  132} (1993) 377--388.

\bibitem{figue1}
J.~M.~A. Figueiredo, A.~Vidal and O.~N. Mesquita {\em J. Crys. Growth\/} {\bf
  166} (1996) 222.

\bibitem{figue2}
J.~M.~A. Figueiredo, M.~Santos, L.~Ladeira and O.~N. Mesquita {\em Phys. Rev.
  Lett.\/} {\bf 71}, 26 (1993) 4397.

\bibitem{CincaBook04}
R.~Gonz\'alez-Cinca, R.~Folch, R.~Ben\'{\i}tez, L.~Ram\'{\i}rez-Piscina,
  J.~Casademunt and A.~Hern\'andez-Machado in E.~Korutcheva and R.~Cuerno,
  eds., {\em Advances in Condensed Matter and Statistical Mechanics\/} (Nova
  Science Publishers, 2004).

\bibitem{Karma96}
A.~Karma and W.~J. Rappel {\em Phys. Rev. E\/} {\bf 53} (1996) R3017.

\bibitem{Karma01}
A.~Karma {\em Phys. Rev. Lett.\/} {\bf 87} (2001) 115701.

\bibitem{almgren-one-sided}
R.~F. Almgren {\em SIAM J. Appl. Math.\/} {\bf 59} (1999) 2086.

\bibitem{Karma99}
A.~Karma and W.-J. Rappel {\em Phys. Rev. E\/} {\bf 60} (1999) 3614.

\bibitem{Pavlik99}
S.~Pavlik and R.~Sekerka {\em Physica A\/} {\bf 268} (1999) 283.

\bibitem{Pavlik00}
S.~Pavlik and R.~Sekerka {\em Physica A\/} {\bf 277} (2000) 415.

\bibitem{Halperin74}
B.~Halperin, P.~Hohenberg and S.~Ma {\em Phys. Rev. B\/} {\bf 10} (1974) 139.

\bibitem{wheeler1}
A.~A. Wheeler, W.~J. Boettinger and G.~B. McFadden {\em Phys.\ Rev. A\/} {\bf
  45} (1992) 7424.

\bibitem{wheeler2}
A.~A. Wheeler, W.~J. Boettinger and G.~B. McFadden {\em Phys.\ Rev. E\/} {\bf
  47} (1993) 1893.

\bibitem{benitez}
R.~Ben\'{\i}tez and L.~Ram\'{\i}rez-Piscina submitted to Phys. Rev. Lett.
  (2004).

\bibitem{mullins-sekerka}
W.~W. Mullins and R.~F. Sekerka {\em J. Appl. Phys.\/} {\bf 35} (1964) 444.

\bibitem{Karma98}
W.~Losert, D.~A. Stillman, H.~Z. Cummins, P.~Kopczy\'nski, W.~J. Rappel and
  A.~Karma {\em Phys. Rev. E\/} {\bf 58} (1998) 7492.

\bibitem{oswald}
P.~Oswald, J.~Bechhoefer and A.~Libchaber {\em Phys. Rev. Lett.\/} {\bf 58}, 22
  (1987) 2318.

\bibitem{simon}
A.~J. Simon, J.~Bechhoefer and A.~Libchaber {\em Phys. Rev. Lett.\/} {\bf 61},
  22 (1988) 2574.

\bibitem{dresden}
R.~Ben\'{\i}tez and L.~Ram\'{\i}rez-Piscina vol.~32 of {\em Lecture Notes in
  Computational Science and Engineering\/} 160--165 (Springer-Verlag, Berlin,
  2003).

\end{thebibliography}


\end{document}